\documentclass[12pt]{article}
\usepackage{amsmath}
\usepackage{graphicx}
\usepackage{enumerate}
\usepackage{url} 

\newcommand{\blind}{1}

\usepackage[normalem]{ulem}

\usepackage{amssymb}
\usepackage{amsmath,amsthm, mathtools,commath}
\usepackage{graphics, color}
\usepackage{booktabs}
\usepackage{arydshln}

\usepackage{graphicx}
\usepackage{epsfig}
\usepackage{makeidx}
\usepackage{kotex}
\usepackage{fullpage}
\usepackage{comment}

\usepackage[round]{natbib}
\newcommand{\T}{\top}

\newcommand{\biblist}{\begin{list}{}
{\listparindent 0.0cm \leftmargin 0.50cm \itemindent -0.50 cm
\labelwidth 0 cm \labelsep 0.50 cm
\usecounter{list}}\clubpenalty4000\widowpenalty4000}
\newcommand{\ebiblist}{\end{list}}

\newcommand{\bx}{\bm{x}}

\newcommand{\bz}{\bm{z}}

\newcommand{\bbeta}{{\mbox{\boldmath$\beta$}}}
\newcommand{\blambda}{{\mbox{\boldmath$\lambda$}}}

\usepackage{xcolor}

\usepackage{graphicx, graphics, color, subcaption}

\usepackage{amsmath, amssymb, amsfonts, amsthm}
\usepackage{mathtools, mathrsfs, bbm, bm, commath}

\newtheorem{proposition}{Proposition}
\newtheorem{theorem}{Theorem}
\newtheorem{remark}{Remark}
\newtheorem{corollary}{Corollary}

\DeclareMathOperator*{\argmax}{arg\,max}
\DeclareMathOperator*{\argmin}{arg\,min}

\usepackage{float, multirow}

\usepackage{comment}

\usepackage{enumitem} 

\usepackage{xr}
\numberwithin{equation}{section}

\begin{document}
\def\spacingset#1{\renewcommand{\baselinestretch}%
{#1}\small\normalsize} 
\spacingset{1}

\if1\blind
{
  \title{\bf 
  Debiased calibration estimation using generalized entropy  in survey sampling  }
  \author{Yonghyun Kwon \\
  Department of Mathematics, Korea Military Academy, Seoul, Republic of Korea \\
  and \\
   Jae Kwang Kim \\
  Department of Statistics, Iowa State University, Iowa, USA\\
  and \\
  Yumou Qiu \\
  School of Mathematical Sciences, Peking University, Beijing, China  
  }
\date{}
\maketitle
} \fi

\if0\blind
{
 \bigskip
 \bigskip
 \bigskip
 \begin{center}
   {\LARGE\bf 
   Debiased calibration estimation using generalized entropy  in survey sampling   }
\end{center}
 \medskip
} \fi

\bigskip




\begin{abstract}
Incorporating auxiliary information into the survey estimation is a fundamental problem in survey sampling. Calibration weighting is a widely used technique to integrate such information by adjusting design weights to meet benchmarking constraints. Traditional methods, such as those proposed by \cite{deville1992}, solve this problem by minimizing a distance between calibrated and design weights. In this paper, we propose a novel calibration framework that instead maximizes a generalized entropy function subject to two constraints: a benchmarking constraint to improve efficiency and a debiasing constraint involving design weights to ensure design consistency. This approach avoids placing design weights in the objective function and instead incorporates them through the constraint structure. We establish the asymptotic properties of the proposed estimator, including design consistency and asymptotic normality, and demonstrate that under Poisson sampling, a specific contrast-entropy function minimizes the asymptotic variance among a broad class of entropy functions. Simulation studies and an empirical application to agricultural survey data illustrate the advantages of our method, particularly in the presence of model misspecification or informative sampling designs. 
We demonstrate a real-life application using agricultural survey data collected from Kynetec, Inc. 
\end{abstract}

\noindent%
{\it Keywords:} Contrast entropy, empirical likelihood, generalized regression estimation,   selection bias.
\vfill

\newpage
\spacingset{1.9} 

\section{Introduction}
\label{ch4sec1}
Probability sampling is a classical tool for securing a representative sample from a target
population. 
Once a probability sample is obtained, researchers can employ design consistent estimation methods, such as Horvitz-Thompson estimation, to estimate the parameters of the population. Statistical inferences, such as confidence intervals, can be justified from the probability sample  using the large sample theory \citep{fuller2009sampling}.

However, the Horvitz-Thompson (HT) estimator is not necessarily efficient as it does not incorporate all available information effectively. Improving the  efficiency of the HT estimator is one of the fundamental problems in survey sampling. A classical approach to improving efficiency involves the use of additional information obtained from external sources, such as census data. To incorporate auxiliary information into the final estimate, the design weights are adjusted to meet the benchmarking constraints imposed by the auxiliary information.  
This weight modification method to satisfy the benchmarking constraint is known as calibration weighting. When the auxiliary variables used for calibration are correlated with the study variable of interest, the resulting calibration estimator is more efficient than the HT estimator.
While calibration weighting also plays a role in mitigating selection bias in nonprobability samples, as highlighted by \cite{dever2016} and \cite{elliott2017inference}, our discussion will focus on its application within probability samples.

The literature on calibration weighting is very extensive. 
\cite{isaki1982survey} used a linear regression superpopulation model to construct the regression calibration weights and showed that the resulting estimator is optimal in the sense that its anticipated variance achieves the lower bound of \cite{godambe1965}.  \cite{deville1992} developed a unified framework for calibration estimation and showed the asymptotic equivalence between the calibration estimator and the generalized regression (GREG) estimator. 
Breidt and Opsomer developed a series of nonparametric regression estimators \citep{breidt05, breidt2017model}
that can be understood as nonparametric calibration weighting. 
\cite{dagdoug2023} employ random forest to develop a nonparametric model calibration. 
See \cite{haziza2017construction, devaud2019} for a comprehensive review of calibration weighting methods in survey sampling.

Most existing calibration weighting methods
are grounded in the framework established by  \cite{deville1992}, which utilizes a distance measure between the design weights and the final weights to address the optimization problem within the calibration constraints. In this paper, we introduce a novel alternative framework based on maximizing a generalized entropy function subject to two sets of constraints: a benchmarking constraint on auxiliary variables to reduce variance, and a \emph{debiasing constraint} involving design weights to ensure design consistency. Unlike traditional approaches, our method does not place design weights in the objective function but instead incorporates them through the constraint structure.

The idea of employing a debiasing constraint within the calibration weighting framework is not entirely new. \cite{Qin2002} is perhaps the first attempt to correct selection bias using empirical likelihood in the context of missing data.  
\cite{berger2016empirical} employed debiasing constraints in empirical likelihood for survey sampling contexts. Chapter 2 of \cite{fuller2009sampling} discussed the incorporation of the debiasing constraint into linear regression models. Building on these ideas, our framework unifies entropy-based calibration with debiasing constraints and provides a general method for constructing efficient and design-consistent estimators.

We show that the resulting calibration estimator is asymptotically equivalent to a debiased prediction estimator based on an augmented regression model that includes the debiasing covariate. Furthermore, the choice of entropy function affects the efficiency of the final estimator. Under Poisson sampling, we identify the \emph{contrast-entropy} function as the optimal choice for minimizing asymptotic variance. The proposed method includes empirical likelihood calibration as a special case and leads to estimators with improved performance, particularly under model misspecification or informative sampling designs.

We also consider a practical extension where design weights are unavailable outside the sample. In such cases, we develop a modified entropy-based method that constructs the calibration weights without using the population mean of the debiasing control variable, preserving key asymptotic properties such as design consistency and asymptotic normality.

The paper is organized as follows. In Section \ref{ch4sec2}, the basic setup and the research problem are introduced. In Section \ref{ch4sec3}, we present the proposed method using a generalized entropy calibration and give an illustration using empirical likelihood as a special case. In Section \ref{ch4sec4}, the asymptotic properties of the proposed method are rigorously derived, and a consistent variance estimator is also presented. 
In Section \ref{ch4sec5}, we show that the optimal entropy function in the context of calibration estimation is achieved with the contrast-entropy function. 
In Section \ref{ch4sec7}, we present a modification of the proposed method when the design weights are not available outside the sample. Results of a limited simulation study are presented in Section \ref{ch4sec8}. In Section \ref{ch4sec9}, we demonstrate a real-life application using
agricultural survey data collected from Kynetec, Inc. Some concluding remarks are made in Section \ref{ch4sec10}. 
All the technical proofs are relegated to the supplementary material (SM).

\section{Basic setup}
\label{ch4sec2}
Consider a finite population $U = \set{1, \cdots, N}$ of $N$ units,  where  $N$ is known.  A sample $A$ of size $n$ is selected from the finite population using a given probability sampling design. Let $\delta_i$ be the sampling indicator variable, where $\delta_i = 1$ if unit $i$ is included in the sample $A$,  and $\delta_i = 0$ otherwise. We assume that first-order inclusion probabilities $\set{\pi_i: i \in U}$ are available throughout the population. We will relax this assumption later in Section \ref{ch4sec7}.

Let $y_i$ denote  the study variable of interest, available only in sample $A$. Our goal is to estimate the finite population total $\theta_{\rm N} = \sum_{i \in U} y_i =  \sum_{i = 1}^N y_i$ using the sample $A$. We consider a class of linear estimators
\begin{equation}
    \hat \theta = \sum_{i \in A} \omega_i y_i, \label{lin}
\end{equation}
where $\omega_i$ is not a function of $\{y_i: i \in A\}$. The Horvitz-Thompson (HT) estimator uses   $\omega_i = \pi_i^{-1}$, yielding a design-unbiased 
estimator of $\theta_{\rm N}$, although it may be  inefficient.

In many practical situations, in addition to the study variable $y_i$, we observe $p$-dimensional auxiliary variables $\bm x_i = (x_{i1}, \ldots, x_{ip})^{\T}$ with known population totals. 
In this case, to incorporate the information of covariates, we often require that the final weights  satisfy  
\begin{equation}
\sum_{i \in A} \omega_i \bm x_i = \sum_{i \in U} \bm x_i.
\label{calib} 
\end{equation}
Constraint (\ref{calib}), often called a calibration or benchmarking constraint, ensures the weighted sum of auxiliary variables in the sample matches known population totals.
Generally speaking, the calibration estimator is more efficient than the HT estimator when the study variable $y$ of interest is related to the auxiliary variable $\bm x$. In particular, if $y_i = \bx_i^{\T} \bbeta$ holds exactly, the calibration estimator achieves zero mean squared error.
Thus, $y_i= \bx_i^{\T} \bbeta + e_i$ can be viewed as a working model for calibration. 
However, the calibration condition itself does not guarantee design consistency.

To achieve the design consistency of the calibration estimator, \cite{deville1992} proposed solving an optimization problem that minimizes the distance 
\begin{equation}
    D(\bm \omega, \bm d) = \sum_{i \in A}d_i G(\omega_i / d_i),  \label{DS}
\end{equation}
which quantifies the discrepancy 
between the calibration weights $\bm \omega = \del{\omega_i}_{i \in A}$ and the design weights $\bm d = \del{\pi_i^{-1}}_{i \in A}$,  subject to the calibration constraints in (\ref{calib}), where $d_i = \pi_i^{-1}$ for $i \in A$ and $G(\cdot): \mathcal V \to \mathbb R$ is a nonnegative function that is strictly convex, differentiable, and $G'(1) = 0$. 
The domain $\mathcal V$ of $G(\cdot)$ is an open interval in $\mathbb R$. The distance measure $D(\bm \omega, \bm d)$ in \eqref{DS} serves as the divergence between two discrete measures $\bm \omega$ and $\bm d$. For example, {$G(v) = v \log(v) - v + 1$}, with domain $\mathcal V \subseteq (0, \infty)$, corresponds to the {reverse} Kullback-Leibler divergence, while {$G(v) = (v - 1)^2$}, with the domain $\mathcal V \subseteq  (-\infty, \infty)$, corresponds to the Chi-squared distance from 1. 

Let $\hat{\omega}_{{\rm ds}, i}$ be the solution to the above optimization problem, and let $\hat{\theta}_{\rm ds} = \sum_{i \in A} \hat{\omega}_{{\rm ds}, i} y_i$ be the Deville and S\"{a}rndal's (DS) estimator. 
Under mild regularity conditions, 
$\hat{\theta}_{\rm  ds}$
is asymptotically equivalent to the generalized regression (GREG) estimator given by 
\begin{equation}
\hat{\theta}_{\rm greg} = \sum_{i \in U}\bm x_i^\T \hat {\bm \beta} + \sum_{i \in A} d_i (y_i - \bm x_i^\T \hat{\bm \beta}) 
\label{greg}
\end{equation}
where 
\begin{equation}
    \hat{\bm \beta} = \bigg(\sum_{i \in A}d_i\bm x_i \bm x_i^\T\bigg)^{-1}\sum_{i \in A}d_i\bm x_i y_i.
    \label{linc}
\end{equation}

Note that the asymptotic expansion $\hat{\theta}_{\rm greg}$ of $\hat{\theta}_{\rm  ds}$ is free of the $G(\cdot)$ function and it is expressed as the sum of two terms; the prediction term and the bias correction term. The bias correction term is calculated from the sample using the HT estimation of the negative bias of the prediction estimator.  
The bias-corrected prediction estimator is also called a debiased prediction estimator in the causal inference literature \citep{athey2018}.  The debiasing property comes from the fact that the objective function in (\ref{DS}) is minimized at $\omega_i=d_i$ if the calibration constraints are satisfied under these design weights, which is nearly true when the sample size is sufficiently large. 
Thus, the final calibration weights should converge to the design weights as the sample size increases. 

Although DS and GREG estimators have the advantage of eliminating bias, they 
may not be efficient
because only $\bx$ is used in the prediction term and the regression coefficient $\hat{\bm \beta}$ is free of the $G(\cdot)$ function.
In the following sections, we propose another approach to the debiased calibration estimator that can be more efficient than the classical calibration estimator that uses (\ref{DS}). The basic idea is to incorporate bias correction into the calibration constraint, which will be called a debiasing constraint.
The benchmarking constraint targets variance reduction through auxiliary information, while the debiasing constraint addresses selection bias and promotes design consistency.
The use of a debiasing constraint in calibration weighting is similar in spirit to interval bias calibration (IBC) in prediction estimation \citep{Firth1998}. The debiasing constraint in the calibration estimator plays the role of IBC in the prediction estimator.

\section{Methodology}
\label{ch4sec3}

Instead of minimizing the weight  distance measure in \eqref{DS}, we now consider maximizing the generalized entropy \citep{gneiting2007strictly} that does not employ the design weights: 

\begin{equation}
    H(\bm \omega) = -\sum_{i \in A} G(\omega_i) , \label{epy}
\end{equation}
where $G(\cdot):\mathcal V \to \mathbb R$ is a strictly convex and differentiable function with an open domain $\mathcal V$ and $\bm \omega$ denotes the vector of $\{\omega_i: i \in A\}$. 
The empirical likelihood is a special case of \eqref{epy} with $G(\omega_i) = -\log \omega_i$,  while the Shannon-entropy uses $G(\omega_i) = \omega_i \log \omega_i$.

To guarantee design consistency, in addition to the benchmarking  calibration constraints in (\ref{calib}), 
we propose including the following design calibration constraint 
\begin{equation}
    \sum_{i \in A}\omega_i g(d_i) = \sum_{i \in U} g(d_i),
    \label{dcc}
\end{equation}
where $g(\omega) = d G(\omega)/ d \omega$ denotes the first-order derivative of $G(\cdot)$. 
The constraint in (\ref{dcc}) is the key constraint to make the proposed calibration estimator design consistent, which is called the \textit{debiasing} calibration constraint. The mapping $d_i \mapsto g(d_i)$ is called the debiasing transformation. Including the debiasing constraint in the generalized entropy function is our main proposal. While the primary reason for including the debiasing constraint is to achieve design consistency, we can also improve the efficiency of the resulting estimator as it incorporates an additional covariate in the working regression model for calibration.

Our goal is to find the calibration weights $\bm \omega$ that maximize the generalized entropy in \eqref{epy} under the calibration constraints in (\ref{calib}) and (\ref{dcc}). 
The optimization problem of interest can be formulated as follows
\begin{equation}
    \hat {\bm \omega} = \argmin_{\omega_i \in \mathcal V} \sum_{i \in A} G(\omega_i) \mbox{ \ subject to \ }  \sum_{i \in A} \omega_i \bm z_i = \sum_{i \in U} \bm z_i, \label{prob}
\end{equation}
where  $\bm z_i^{\T} = (\bm x_i^{\T}, g(d_i)) \in \mathbb R^{p + 1}$ is the augmented covariate vector including both auxiliary information and the debiasing transformation. Note that $\hat{\bm \omega}$ is the vector of $\{\hat{\omega}_i: i \in A\}$. Let  $\Omega_A = \big\{\bm \omega = (\omega_i)_{i \in A}: \sum_{i \in A} \omega_i \bm z_i = \sum_{i \in U} \bm z_i \mbox{ \ and \ } \omega_i \in \mathcal V \big\}$.
If $\Omega_A$ is nonempty, a solution $\hat {\bm \omega}$ to (\ref{prob}) exists, and the proposed generalized entropy calibration (GEC) estimator of the population total $\theta_{\rm N}$ is constructed as $\hat{\theta}_{\rm gec} = \sum_{i \in A} \hat{\omega}_i y_i$.

The Karush-Kuhn-Tucker (KKT) conditions for the primal problem in (\ref{prob}) are
\begin{equation}
g(\omega_i) - \blambda^{\T} \bz_i = 0 \mbox{ \ and \ } \sum_{i \in A} \omega_i \bm z_i = \sum_{i \in U} \bm z_i,
\label{eq:KKT}\end{equation}
where $\blambda = (\blambda_1^{\T}, \lambda_2)^{\T}$ are the Lagrange multipliers for the dual problem of (\ref{prob}). 
Since $G$ is strictly convex, Slater's condition is satisfied and strong duality holds from Section 5 of \cite{boyd2004convex}, and the solution to (\ref{prob}), if exists, is unique.
Then, $\hat {\bm \omega}$ is the solution to (\ref{prob}) if and only if 
\begin{equation}
\hat \omega_i = \hat \omega_i(\hat{\bm \lambda}) 
= g^{-1}\{\hat{\bm \lambda}_1^\T  \bm x_i + \hat{\lambda}_2 g (d_i) \},
\label{fwgt}
\end{equation}
and $\hat{\bm \lambda} = (\hat{\bm \lambda}_1^\T, \hat{\lambda}_2)^{\T}$ satisfy the KKT conditions in (\ref{eq:KKT}). 
Let $g(\mathcal V) = \{ g(\omega): \omega \in \mathcal V\}$ and $\Lambda_A = \big\{\bm \lambda: \bm \lambda^\T \bm z_i \in g(\mathcal V) \mbox{ \ for all \ } i \in A \big\}$. Let  $F(u) = - G(g^{-1}(u)) + g^{-1}(u) u$ for $u \in g(\mathcal V)$  be the convex conjugate function of $G(\omega)$, and $f(u) = d F(u)/ d u$ be the first-order derivative of $F(\omega)$. By the chain rule, we can obtain  $f(u) =  g^{-1}(u)$ for $u \in g(\mathcal V)$. 
It is shown in the SM that $\Lambda_A$ is non-empty and the solution to (\ref{eq:KKT}) can be obtained by the optimization problem
\begin{equation}
    \hat{\bm \lambda} = \argmin_{\bm \lambda \in \Lambda_A} \sum_{i \in A} F(\bm \lambda^\T \bm z_i) - \bm \lambda^\T \sum_{i \in U} \bm z_i
    \label{dual}
\end{equation} 
as $n, N \to \infty$. 
Thanks to the existence of the solution to \eqref{dual} with probability approaching to 1 ($w.p.a.1$), this implies that $\Omega_A$ is non-empty and the solution to \eqref{prob} exists $w.p.a.1$. Solving the dual problem in \eqref{dual}, which involves only $p+1$ parameters, offers computational advantages over the $n$-dimensional primal problem of \eqref{prob}.

To understand the role of the debiasing constraint in (\ref{dcc}), first note that, to make the proposed calibration estimator $\hat{\theta}_{\rm gec}$ design consistent, we require the calibration weight $\lim \hat{\omega}_i \to d_i$ as $n, N \to \infty$. 
If the constraint $\sum_{i \in A}\omega_i h(d_i) = \sum_{i \in U} h(d_i)$ is imposed for another function $h(d_i)$ instead of (\ref{dcc}), we obtain the calibration weight 
$$ \hat{\omega}_i = g^{-1}( \hat{\blambda}_1^\top \bx_i + \hat{\lambda}_2 h(d_i) )$$
as the solution to the generalized entropy optimization problem in (\ref{prob}) with $g(d_i)$ replaced by $h(d_i)$. 
Let $\tilde{\bm \lambda}_1$ and $\tilde{\lambda}_2$ be the probability limits of $\hat{\blambda}_1$ and $\hat{\lambda}_2$ as $n, N \to \infty$, and $\tilde{\omega}_i = g^{-1}( \tilde{\blambda}_1^\top \bx_i + \tilde{\lambda}_2 h(d_i) )$.
To achieve design consistency, we need $\tilde{\omega}_i = d_i$ for all $i$, which is equivalent to $\tilde{\blambda}_1^\top \bx_i + \tilde{\lambda}_2 h(d_i) = g(d_i)$ for all $i$.
Therefore, we must have 
$\tilde{\blambda}_1 = \mathbf{0}$, $\tilde{\lambda}_2 = 1 / c$ and $h(d_i) = c g(d_i)$ for a constant $c \neq 0$. 
Other choices of $h(d_i)$ that are not proportional to $g(d_i)$ cannot achieve design consistency.

\cite{deville1992} showed that the calibration weights $\hat{\omega}_{{\rm  ds}, i}$ using the divergence measure in (\ref{DS}),  
can be expressed as 
\begin{equation}
\hat{\omega}_{{\rm  ds}, i} = \hat{\omega}_{{\rm  ds}, i}(\hat{\bm \lambda}_{\rm  ds, 1}) = d_i g^{-1} ( \hat{\bm \lambda}_{\rm  ds, 1}^\T \bm x_i ) , 
\label{fwgt0}
\end{equation}
where $\hat{\bm \lambda}_{\rm  ds, 1}$ is the Lagrange multiplier   satisfying  the calibration constraints in (\ref{calib}). 
They also showed that $\hat{\bm \lambda}_{\rm  ds, 1} \to {\bm 0}$ in probability as $n, N \to \infty$ under mild regularity conditions, which implies $\hat{\omega}_{{\rm  ds}, i} /d_i \rightarrow 1$.  By comparing  (\ref{fwgt}) with (\ref{fwgt0}), we can see that
the main distinction lies  in the way that the selection probabilities $\{d_i\}$ are utilized. 
The calibration method of \cite{deville1992} uses $\{d_i\}$ in the objective function. On the other hand, 
the proposed method uses $\{d_i\}$  through the debiasing calibration constraint in \eqref{dcc}. An additional parameter $\lambda_2$ is introduced to reflect the debiasing constraint in \eqref{dcc}.  

It is worth noticing that the constraint in (\ref{dcc}) can be applied to the DS approach as well, which leads to the augmented DS weight $\hat{\omega}_{{\rm  ads}, i} = d_i g^{-1} ( \hat{\bm \lambda}_{\rm  ads, 1}^\T \bm x_i + \hat{\lambda}_{\rm  ads, 2} g(d_i) )$, where $\hat{\bm \lambda}_{\rm  ads, 1}$ and $\hat{\lambda}_{\rm  ads, 2}$ satisfy the constraints in (\ref{calib}) and (\ref{dcc}). The augmented DS estimator is constructed as $\hat{\theta}_{\rm ads} = \sum_{i \in A} \hat{\omega}_{{\rm ads}, i} y_i$, which uses the same set of calibration constraints as the proposed estimator $\hat{\theta}_{\rm gec}$. In the following two sections, we provide a theoretical comparison for $\hat{\theta}_{\rm gec}$ and $\hat{\theta}_{\rm ads}$, and show the GEC estimator with the contrast entropy is more efficient than the corresponding augmented DS estimator. 
We also extend the proposed GEC estimator to the case where the design weights $\{d_i: i \not\in A\}$ are not available by modifying the objective function in (\ref{prob}) and dropping the debiasing constraint in (\ref{dcc}). Under this case, we compare the efficiency of the modified GEC estimator with the original DS estimator $\hat{\theta}_{\rm ds}$.
See Section \ref{ch4sec7} for details.

The following provides examples of entropy functions for the proposed method.

\noindent{\bf Empirical Likelihood Example.}
The empirical likelihood  objective function of (\ref{epy}) uses $G(\omega)= - \log (\omega)$. 
Since $g(\omega) = - \omega^{-1}$ in this case, the debiasing constraint in (\ref{dcc}) takes the form
$\sum_{i \in A} \omega_i \pi_i = \sum_{i \in U} \pi_i = {\rm E}(n)$. When the sample size $n$ is fixed, $\sum_{i \in U}\pi_i$ is equal to $n$.
The proposed calibration method using the empirical likelihood objective function solves the optimization problem
\begin{equation}
\hat {\bm \omega} = \argmax_{\omega_i > 0} \ell^{\rm EL}(\bm \omega)
= \argmax_{\omega_i > 0} \sum_{i \in A} \log \omega_i \mbox{ \ subject to \ }
\sum_{i \in A} \omega_i \pi_i = \sum_{i \in U} \pi_i \mbox{ \ and \ } (\ref{calib}). 
\label{el}
\end{equation}
The logarithm in $\log (\omega_i)$ ensures that $\omega_i$ are all positive.
The use of \eqref{el} for complex survey design was considered by \cite{berger2016empirical}.
For comparison, the empirical likelihood weight of \cite{deville1992} is obtained by minimizing the Kullback-Leibler divergence $D(\bm w, \bm d) = \sum_{i \in A}d_i \log \del{d_i / \omega_i}$, which  
is equivalent to maximizing the pseudo empirical likelihood (PEL) proposed by \cite{chen1999pseudo}:
\begin{equation}
    \ell^{\rm PEL}(\bm \omega) = \sum_{i \in A} d_i \log (\omega_i) \label{pel}
\end{equation}
subject to the calibration constraint in (\ref{calib}).  Other examples of generalized entropies and their debiasing transformation function $g(d_i)$ can be found in Table \ref{tab1tmp}. 

\begin{table}[ht]
\small
\centering
\begin{tabular}{ccccc}
\hline\hline
Entropy & $G(\omega)$ & $g_i = g(d_i)$ & $1 / g'(d_i)$ & Domain $\mathcal V$ \\ \hline
Squared loss & $\omega^2 / 2$ & $d_i$ & $1$ & $(-\infty, \infty)$ \\
Empirical likelihood & $-\log \omega$ & $-d_i^{-1}$ & $d_i^{2}$ & $(0, \infty)$ \\
Exponential tilting & $\omega \log(\omega) - \omega$ & $\log d_i$ & $d_i$ & $(0, \infty)$ \\
Shifted Exp tilting & $(\omega - 1)\log(\omega - 1) - \omega$ & $\log(d_i - 1)$ & $d_i - 1$ & $(1, \infty)$\\
Contrast entropy & 
$(\omega - 1)\log(\omega - 1) - \omega \log(\omega)$
& $\log(1 - d_i^{-1})$ & $d_i^2 - d_i$ & $(1, \infty)$       \\
%
Pseudo-Huber & $M^2 \{1 + \del{\omega/M}^2\}^{1/2}$ & $d_i \{1 + (d_i / M)^2\}^{-1/2}$ & $(d_i / g_i)^3$ & $(-\infty, \infty)$\\
Hellinger distance & $-4\omega^{1/2}$ & $-2d_i^{-1/2}$       & $d_i^{3 / 2}$ & $(0, \infty)$ \\
Inverse & $1 / (2\omega)$ & $-d_i^{-2} / 2$ & $d_i^{3}$ & $(0, \infty)$ \\
R\'enyi entropy & $r^{-1}(r + 1)^{-1}\omega^{r + 1}$ & $r^{-1} d_i^{r}$ & $ d_i^{-r + 1}$ & $(0, \infty)$ \\ 
\hline\hline
\end{tabular}
\caption{Examples of generalized entropies with the corresponding $G(\omega)$, debiasing transformation function  $g_i = g(d_i) = g(\pi_i^{-1})$ and the regression weight $1 / g'(d_i)$ in (\ref{gammahat}), where $g'(d_i)$ is the first-order derivative of $g(d_i)$ with respect to $d_i$ and the R\'enyi entropy requires $r \neq 0, -1$.} 
\label{tab1tmp}
\end{table}

Let $n_0 = {\mathbb E}(n) = \sum_{i \in U} \pi_i$. Under the case of $\pi_i \rightarrow 0$ and $n_0 = o(N)$ in the asymptotic setup, the sample size $n$ would be much smaller than $N$ and $d_i \to \infty$ which makes $g(d_i)$ a trivial value for all $i$ asymptotically. To cover this case, we use a scaled weight $d_i^*= n_0 d_i/N$ in the augmented calibration and modify the proposed approach in (\ref{prob}) as
\begin{equation}
\hat {\bm \omega} = \argmin_{ n_0 \omega_i / N \in \mathcal V} \sum_{i \in A} G \bigg( \frac{n_0}{N}\omega_i \bigg) \mbox{ \ subject to \ }  \sum_{i \in A} \omega_i \bm z_i = \sum_{i \in U} \bm z_i, \label{Prob2}
\end{equation}
where $\bm z_i^{ \T} = (\bm x_i^{\T}, g_i^{\ast})$ and $g_i^{\ast} = g( d_i^*)$. Note that the unscaled calibration weighting method in (\ref{prob}) is a special case of the scaled calibration weighting method in (\ref{Prob2}). To see this, we can define $\tilde{G}(\omega) = G(n_0 \omega / N )$ if $n_0$ and $N$ are of the same order, then (\ref{Prob2}) becomes (\ref{prob}). Therefore, we do not distinguish the notation $\bm z_i$ for the unscaled and scaled approaches.

\section{Statistical properties}
\label{ch4sec4}

In this section, we establish the asymptotic properties of the proposed generalized entropy calibration (GEC) estimator. 
To facilitate a unified asymptotic analysis, we  consider an increasing sequence of finite populations and samples as in \cite{isaki1982survey} and 
present the results under the scaled calibration framework introduced in (\ref{Prob2}), which covers the case of unscaled weighting in (\ref{prob}) if $n_0$ and $N$ are of the same order.

We first state a set of regularity conditions that ensure the asymptotic validity of the proposed method.

\begin{enumerate}[label={[A\arabic*]}]

\item \label{a1} 
The function $G: \mathcal{V} \to \mathbb{R}$ is strictly convex and continuously differentiable, with $G''(\omega) > 0$ for all $\omega > 0$.
   \item \label{a2}    There exist positive constants $c_1, c_2 \in \mathcal{V}$ such that 
$
c_1 < N \pi_i/n_0 < c_2
$
for $i = 1, \ldots, N$.
\item \label{a3} Let $\pi_{ij}$ be the joint inclusion probability of units $i$ and $j$ and $\Delta_{ij} = \pi_{ij} - \pi_i \pi_j$.
Assume
$$\limsup_{N \to \infty} N^2n_0^{-1} \max_{i, j \in U: i \neq j} \abs{\Delta_{ij}} < \infty.$$

\item \label{a4} 
Assume $\bm \Sigma_{\bm z} = \underset{N \to \infty}{\lim} \sum_{i \in U} \bm z_i {\bm z_i}^{\T} / N$ exists and positive definite, the average 4th moment of $(y_i, \bm x_i^\T)$ is finite such that $\underset{N \to \infty}{\limsup} \sum_{i = 1}^N \norm{(y_i, \bm x_i^\T)}^4 / N < \infty$, and $\bm \Gamma(\bm \lambda) = \underset{N \to \infty}{\lim} \sum_{i \in U} f^\prime(\bm \lambda^\T \bm z_i) \bm z_i {\bm z_i}^{\T} / N$ exists in a neighborhood around $\bm \lambda_0= (\bm\lambda_{10}^{\T}, \lambda_{20})^\T$, where $\bm\lambda_{10} = \bm 0$ and $\lambda_{20} = 1$. 

\end{enumerate}

Following \cite{gneiting2007strictly}, we consider a generalized entropy that is strictly convex. Condition \ref{a1} implies that the solution to the optimization problem \eqref{prob} is unique when it exists. It also indicates that $f(\cdot) = g^{-1}(\cdot)$ is differentiable \citep{deville1992}.
Condition \ref{a2}  avoids {extremely large  weights which may cause instability in estimation} 
and prevents the random sample $A$ from being concentrated on a few units in the population. 
Condition \ref{a3} ensures that the mutual dependence of the two sampling units is not too strong, which is satisfied under many classical survey designs \citep{robinson1983asymptotic, breidt2000local}.
We also assume that the population has a finite average fourth moment, and the covariates $\bm z$'s are asymptotically of full rank in Condition \ref{a4}.
Note from (\ref{fwgt}) that
$f^\prime(\bm \lambda_0^\T \bm z_i) = 1 / g^\prime(d_i^*)$ and $\bm \Gamma(\bm \lambda_0) = \underset{N \to \infty}{\lim} \sum_{i \in U} \{ g^\prime(d_i^*) \}^{-1} \bm z_i {\bm z_i}^{\T} / N$, which is finite from Conditions \ref{a1} and \ref{a2} and the existence of $\bm \Sigma_{\bm z}$. Condition \ref{a4} further assumes that $\bm \Gamma(\bm \lambda)$ is finite in a neighborhood of $\bm \lambda_0$.


The following theorem presents the main asymptotic properties of the proposed entropy calibration estimator under standard conditions in survey sampling.

\begin{theorem}[Design consistency]
\label{MainThm}
Suppose Conditions \ref{a1}--\ref{a4} hold. Then, the solution $\hat {\bm \omega}$ to (\ref{Prob2}) exists and is unique with probability approaching to 1.
Furthermore, the proposed entropy calibration estimator $\hat \theta_{\rm gec}= \sum_{i \in A} \hat{\omega}_i y_i$ satisfies 
\begin{equation}
\hat \theta_{\rm gec}  =  \hat \theta_{\rm gec, \ell} + o_p(n_0^{-1/2} N),   \label{cal}
\end{equation}
where 
\begin{equation*}
    \hat \theta_{\rm gec, \ell}  =\sum_{i \in U} {\bm z}_i^\T \bm \gamma_g + \sum_{i \in A} d_i (y_i - {\bm z}_i^\T \bm \gamma_g) 
    \end{equation*}
and
\begin{equation}
\bm \gamma_g = \bigg\{ \sum_{i \in U} \frac{\pi_i \bm z_i \bm z_i^{\T}}{g^\prime(d_i^*)} \bigg\}^{-1} \sum_{i \in U} \frac{\pi_i \bm z_i y_i}{g^\prime(d_i^*)}.
\label{gammahat}
\end{equation}
\end{theorem}


This expansion indicates that the GEC estimator is asymptotically linear and design-consistent, with the bias diminishing at a parametric rate. 
The subscript $g$ in $\bm \gamma_g$ in (\ref{gammahat}) is employed to emphasize its dependency on the generalized entropy function $G(\cdot)$. 

To draw a comparison, we first note that the Deville and S\"{a}rndal's (DS) estimator $\hat \theta_{\rm ds}$ is asymptotically equivalent to the GREG estimator under certain conditions. That is, 
\begin{equation}
\hat \theta_{\rm ds} = \hat \theta_{\rm ds, \ell} + o_p (n_0^{-1/2} N) \mbox{ \ for \ }
\hat \theta_{\rm ds, \ell} = \sum_{i \in U} {\bm x}_i^\T {\bm \beta}_N + \sum_{i \in A} d_i  (y_i - {\bm x}_i^\T {\bm \beta}_N ), 
\label{pel2}
\end{equation}
where ${\bm \beta}_N = \big(\sum_{i \in U} \bm x_i \bm x_i^\T\big)^{-1} \big(\sum_{i \in U} \bm x_i y_i\big)$ is the probability limit of $\hat{\bm \beta}$ used in $\hat{\theta}_{\rm greg}$. The same result also holds for the pseudo-empirical likelihood estimator 
\citep{chen1999pseudo, wurao2006}.
For the augmented DS estimator $\hat \theta_{\rm ads}$ with the additional debiasing constraint introduced after (\ref{fwgt0}), we can establish 
\begin{equation}
\hat \theta_{\rm ads} = \hat \theta_{\rm ads, \ell} + o_p (n_0^{-1/2} N) \mbox{ \ for \ }
\hat \theta_{\rm ads, \ell} = 
\sum_{i \in U} {\bm z}_i^\T {\bm \gamma}_g^{(\rm ads)} + \sum_{i \in A} d_i  (y_i - {\bm z}_i^\T {\bm \gamma}_g^{(\rm ads)} ),
\label{pel2-augment}
\end{equation}
where ${\bm \gamma}_g^{(\rm ads)} = \big(\sum_{i \in U} \bm z_i \bm z_i^\T\big)^{-1} \big(\sum_{i \in U} \bm z_i y_i\big)$ is the probability limit of the regression coefficient $\hat{\bm \gamma}_g^{(\rm ads)} = \big(\sum_{i \in A}d_i\bm z_i \bm z_i^\T\big)^{-1} \big(\sum_{i \in A}d_i\bm z_i y_i \big)$. 

Comparing (\ref{cal}) with (\ref{pel2}), the GEC and DS estimators are asymptotically equivalent if the sampling is non-informative and the working outcome regression model is correctly specified as stated in the following corollary.

\begin{corollary}
\label{corollary1}
Under the conditions of Theorem \ref{MainThm}, if the sampling weight $\pi_i$ is only determined by $\bx_i$ and the data follow a superpopulation linear regression model such that $y_i = \bx_i^{\T} \bbeta_0 + e_i$ with $\mathbb{E}(e_i) = 0$ and $\bx_i$ is independent of $e_i$ $(\bx_i \perp e_i)$, we have $\hat \theta_{\rm gec} = \hat \theta_{\rm ds} + o_p(n_0^{-1/2} N)$ for any entropy function $G(\cdot)$.
\end{corollary}

From the above discussion, we can see that the proposed GEC estimator could be more efficient than the DS and pseudo empirical likelihood estimators as the auxiliary variables $\bm z_i =\del{\bm x_i^{\T}, g(d_i^*)}^{\T}$ use an augmented regression model with an additional covariate $g(d_i^*)$. 
The efficiency gain with the additional covariate will be significant if the sampling design is informative in the sense of \cite{PS2009}, where the additional covariate $g(d_i^*)$ may improve the prediction of $y_i$ as the design weight is correlated with $y_i$ after controlling on $\bm x_i$. 

Comparing the proposed estimator $\hat \theta_{\rm gec}$ to the augmented DS estimator $\hat \theta_{\rm ads}$, both estimators use the same set of covariates. However, the regression coefficient ${\bm \gamma}_g^{(\rm ads)}$ in the asymptotic expansion of $\hat \theta_{\rm ads}$ uses a uniform weight of $1$, while the coefficient $\bm \gamma_g$ in $\hat \theta_{\rm gec}$ uses a weight depending on the choice of $G(\cdot)$. This allows us to improve the efficiency of the proposed method by selecting $G(\cdot)$, as discussed in Section \ref{ch4sec5}.


In order to construct a variance estimator and develop the asymptotic normality of the GEC estimator, we need the following additional conditions.
\begin{enumerate}[label={[B\arabic*]}]
    \item \label{b1} The limit of the design covariance matrix of the HT estimator
    $$
    \bm \Sigma := \lim_{N \to \infty}\frac{n_0}{N^2}\sum_{i, j \in U}\frac{\pi_{ij} - \pi_i \pi_j}{\pi_i \pi_j}\begin{pmatrix}
        y_i y_j & y_i\bm z_j^{\T} \\
        \bm z_i y_j & \bm z_i \bm z_j^{\T}
    \end{pmatrix}
    $$
    exists and is positive-deﬁnite.

\item \label{b2} For any $\set{\varepsilon_i: i \in U}$ with $\limsup_{N \to \infty} N^{-1}\sum_{i \in U}{\varepsilon_i}^4 < \infty$, the HT estimator of $\sum_{i \in U} \varepsilon_i$, $\sum_{i \in A} d_i \varepsilon_i$, is asymptotically normal in the sense that 
$$
\bigg[\mathbb V\bigg\{\sum_{i \in U}\bigg(1 - \frac{\delta_i}{\pi_i}\bigg) \varepsilon_i\bigg\}\bigg]^{-1/2} \sum_{i \in U}\bigg(1 - \frac{\delta_i}{\pi_i}\bigg) \varepsilon_i \stackrel{d}{\to} N(0, 1)
$$
under the sampling design, if
$$
\frac{n_0}{N^2}  
\mathbb V\bigg\{\sum_{i \in U}\bigg(1 - \frac{\delta_i}{\pi_i}\bigg) \varepsilon_i \bigg\} = 
\frac{n_0}{N^2}  \sum_{i,j \in U}\frac{\pi_{ij} - \pi_i \pi_j}{\pi_i \pi_j} \varepsilon_i \varepsilon_j
$$
is positive as $N \to \infty$, where $\stackrel{d}{\to}$ stands for convergence in distribution.
\end{enumerate}

Conditions \ref{b1} and \ref{b2} are standard conditions for survey sampling, which hold in many classical survey designs, including simple random sampling and stratified sampling \citep[Chapter~1]{fuller2009sampling}. Under such conditions, the asymptotic normality of the entropy calibration estimator $\hat{\theta}_{\rm gec}$ can be established.

\begin{theorem}[Asymptotic normality] \label{tm2}
    Under Conditions \ref{a1}--\ref{a4}, \ref{b1} and \ref{b2}, we have
    $$
    \mathbb V^{-1/2}(\hat \theta_{\rm gec}) (\hat \theta_{\rm gec} - \theta_{\rm N}) \stackrel{d}{\to} N\del{0, 1}
    $$
     where $ \mathbb V(\hat \theta_{\rm gec}) =  \mathbb V(\hat \theta_{\rm gec, \ell}) \{1 + o(1)\}$,     $$
    \mathbb V(\hat \theta_{\rm gec, \ell}) = \sum_{i,j \in U}(\pi_{ij} - \pi_i \pi_j)\frac{(y_i - {\bm z}_i^\T \bm \gamma_g)}{\pi_i} \frac{(y_j - {\bm z}_j^\T \bm \gamma_g)}{\pi_j}
    $$
    and $\bm \gamma_g$ is defined in (\ref{gammahat}). 
\end{theorem}

Theorem \ref{tm2} implies that the design variance of $\hat \theta_{\rm gec}$ depends on the prediction error of the regression of $y$ on $\bm z$. 
It suggests that the proposed estimator will perform better if the debiased calibration covariate $g(d_i^*)$ contains additional information on predicting $y$. 
By Theorem \ref{tm2}, 
the variance of $\hat \theta_{\rm gec}$ can be estimated by 
\begin{equation}
    \hat{\mathbb V}(\hat \theta_{\rm gec}) = \sum_{i,j \in A}\frac{(\pi_{ij} - \pi_i \pi_j)}{\pi_{ij}}\frac{(y_i - {\bm z}^{\T}_i \hat{\bm \gamma}_g)}{\pi_i}\frac{(y_j - \bm z^{\T}_j \hat{\bm \gamma}_g)}{\pi_j},\label{varhat}
\end{equation}
where $\hat{\bm \gamma}_g = \big\{\sum_{i \in A} \bm z_i \bm z_i^{\T} / g'(d_i^*) \big\}^{-1} \big\{\sum_{i \in A} \bm z_i y_i / g'(d_i^*) \big\}$.
It is shown in the SM that the ratio 
$\hat {\mathbb V}(\hat \theta_{\rm gec}) / {\mathbb V}(\hat \theta_{\rm gec}) \stackrel{p}{\to} 1$ as $n_0, N \to \infty$ under the conditions of Theorem \ref{tm2} and a technical condition [A5] in the SM that regulates the dependence in high-order inclusion probabilities. 

The asymptotic results in Theorems \ref{MainThm} and \ref{tm2} are valid for any choice of the entropy function $G(\cdot)$. In the next section, we discuss the optimal choice of $G(\cdot)$ in the Poisson sampling design and compare the proposed estimator with the augmented DS estimator, where $\{\delta_i\}_{i \in U}$ are independent. 
To simplify notation, in the following two sections, we only consider the case where $n_0$ and $N$ are the same order and use the unscaled weights in (\ref{prob}). The same result applies to the general case with the scaled weights.

\section{Optimal entropy under Poisson sampling}
\label{ch4sec5}

In this section, we investigate the impact of the choice of entropy function on the efficiency of the proposed estimator.
According to Theorem \ref{MainThm}, the entropy function $G(\cdot)$ influences the proposed estimator $\hat \theta_{\rm gec}$ through the augmented covariates $\bz_i$ and $g^\prime(\cdot)$ in the weights of the regression coefficient $\bm \gamma_g$. As different entropy functions lead to different $\bz_i$, to obtain a design-optimal entropy function without the assumption of the superpopulation outcome regression, we consider a pairwise comparison of two GEC estimators under the same calibration constraints and determine the optimal weight for bias-corrected prediction estimators. 
Namely, we compare the efficiency of two GEC estimators with entropies $G_1(\omega)$ and $G_2(\omega)$, where the calibration constraints include (\ref{calib}) and both $g_1(d_i)$ and $g_2(d_i)$. We also compare the efficiency of the GEC estimator with the augmented DS estimator.

Let $\mathcal{B} = \{\sum_{i \in U} {\bm x}_i^\T \bm \xi + \sum_{i \in A} \pi_i^{-1} (y_i - {\bm x}_i^\T \bm \xi): \bm \xi \in \mathbb{R}^{p}\}$ be a class of bias-corrected prediction estimators for $\theta_{\rm N}$. We do not propose to use an estimator from $\mathcal{B}$ but rather explain the rationale for constructing the optimal entropy function. 
Let $\mathbb{C}$ denote the covariance operator and ${\bm \beta}_{\rm opt} = \big\{ \mathbb{V} \big(\sum_{i \in A} d_i {\bm x}_i \big) \big\}^{-1}
\mathbb{C} \big(\sum_{i \in A} d_i {\bm x}_i,\sum_{i \in A} d_i y_i\big)$. Then, 
\begin{equation}
\hat \theta_{\rm opt} =\sum_{i \in U} {\bm x}_i^\T \bm \beta_{\rm opt} + \sum_{i \in A}\frac{1}{\pi_i} (y_i - {\bm x}_i^\T \bm \beta_{\rm opt})
\label{opt}
\end{equation}
is the design-optimal regression esitmator of $\theta_{\rm N}$ with the smallest variance in the class $\mathcal{B}$  
\citep{montanari1987, rao1994, berger2003towards}. It can also be interpreted as the projection of the HT estimator onto the orthogonal complement of the augmentation space generated by $\sum_{i \in A} d_i {\bm x}_i$ \citep{tsiatis2006}. 
Under Poisson sampling, $\{\delta_i\}_{i \in U}$ are mutually independent, such that $\pi_{ij} = \pi_i \pi_j$ for all $i \neq j$. In this case, 
${\bm \beta}_{\rm opt} = \big( \sum_{i \in U} \pi_i q_i {\bm x}_i {\bm x}_i^{\T} \big)^{-1} \sum_{i \in U} \pi_i q_i {\bm x}_i y_i$, where $q_i = \pi_i^{-2} - \pi_i^{-1}$.
We wish to find the optimal choice of the generalized entropy function $G(\cdot)$ such that the asymptotic expansion of the calibration estimator $\hat \theta_{\rm gec}$ has the same form as the design-optimal regression estimator $\hat \theta_{\rm opt}$ under Poisson sampling. Namely, the regression coefficient $\bm \gamma_g$ of $\hat \theta_{\rm gec}$ has the same form as ${\bm \beta}_{\rm opt}$.

%
To achieve this goal, by Theorem \ref{MainThm}, we only need to find a special entropy function $G(\cdot)$ such that $1/g'(d_i) = d_i^2- d_i,$
which is satisfied with
$g(\omega) =  \log(\omega - 1) - \log(\omega)$.
Thus, the optimal entropy function is 
\begin{equation}
G_{\rm ce}(\omega) = (\omega - 1) \log(\omega - 1) - \omega \log (\omega) \label{crossentropy}
\end{equation}
for $\omega > 1$, which is called the contrast entropy between $(\omega - 1) \log(\omega - 1)$ and $\omega \log (\omega)$. 
Note that the empirical likelihood (EL) and exponential tilting (ET) approaches \citep{kim2010exponential, hainmueller2012entropy} correspond to $G(\omega)$ being $-\log(\omega)$ and $\omega \log(\omega)$, respectively, and choosing $G(\omega) = (\omega - 1) \log(\omega - 1)$ implies a logistic regression model for the inclusion probability $\pi_i$. 
In this view, the optimal entropy function in (\ref{crossentropy})  can be regarded as a contrast between the logistic model and the exponential tilting model for the propensity scores. 

Note that $G_{\rm ce}(\omega)$ in (\ref{crossentropy}) is strictly convex with a negative first derivative $g_{\rm ce}(\omega) 
= \log (1 - \omega^{-1})$ and a positive second derivative $g_{\rm ce}^{\prime}(\omega) = \{\omega (\omega - 1)\}^{-1}$ for $\omega > 1$. It takes negative values for $\omega > 1$ with $\lim_{\omega \downarrow 1} G_{\rm ce}(\omega) = 0$ and $\lim_{\omega \to \infty} G_{\rm ce}(\omega) = -\infty$. The proposed contrast entropy calibration method can be described as the following constrained optimization problem
\begin{eqnarray}
\hat{\bm \omega}_{\rm ce} &=& \underset{\omega_i > 1}{\operatorname{argmin}} \sum_{i \in A} \big\{
(\omega_i - 1) \log(\omega_i - 1) - \omega_i  \log (\omega_i) \big\} \mbox{ \ subject to (\ref{calib}) and } \label{eq:GEL} \\
&& \sum_{i \in A}   \omega_i \log(1 - \pi_i)  = \sum_{i \in U}\log(1 - \pi_i), \label{eq:GEL-IBC} 
\end{eqnarray}
where  
(\ref{eq:GEL-IBC}) is the debiasing calibration constraint, specifically designed for the contrast entropy loss in (\ref{crossentropy}), and (\ref{calib}) is the benchmarking calibration constraint for covariates.

By Theorem \ref{MainThm}, the GEC $\hat \theta_{\rm ce} = \sum_{i \in A} \hat{\omega}_{{\rm ce}, i}y_i$ using the contrast entropy is asymptotically equivalent to the design-optimal regression estimator in (\ref{opt}) with covariate $\bx_i$ and $g_{\rm ce}(d_i)$ under Poisson sampling. 
However, as the additional covariate $g(d_i)$ depends on $G(\omega)$, this result can not be used to compare two GEC estimators. 
To overcome this issue, we introduce the pairwise comparison scheme. Let $\hat \theta_{\rm gec}^{(\rm g1)}$ and $\hat \theta_{\rm gec}^{(\rm g2)}$ be the GEC estimators with entropies $G_1(\omega)$ and $G_2(\omega)$ under the constraints in (\ref{calib}), $\sum_{i \in A}\omega_i g_1(d_i) = \sum_{i \in U} g_1(d_i)$ and $\sum_{i \in A}\omega_i g_2(d_i) = \sum_{i \in U} g_2(d_i)$. Note that the calibration functions $g_1(d_i)$ and $g_2(d_i)$ are added to $\hat \theta_{\rm gec}^{(\rm g2)}$ and $\hat \theta_{\rm gec}^{(\rm g1)}$ so that they have the same set of constraints. 


The following proposition shows the pairwise optimality of the GEC estimator $\hat \theta_{\rm gec}^{(\rm ce)}$ using the contrast entropy $G_{\rm ce}(\omega)$ compared to the GEC estimator $\hat \theta_{\rm gec}^{(\rm g)}$ using any other entropy function $G(\omega)$ under the calibration constraints (\ref{calib}), (\ref{dcc}) and (\ref{eq:GEL-IBC}). Let $\tilde{\bm z}_i = (\bx_i^{\T}, g_{\rm ce}(d_i), g(d_i))^{\T}$.

\begin{proposition}
\label{corollary2}
Under the conditions of Theorem \ref{MainThm} and $n_0$ and $N$ being at the same order, we have $\hat \theta_{\rm gec}^{(\rm ce)} = \hat \theta_{\rm gec, \ell}^{(\rm ce)} + o_p(n_0^{-1/2} N)$ and $\hat \theta_{\rm gec}^{(\rm g)} = \hat \theta_{\rm gec, \ell}^{(\rm g)} + o_p(n_0^{-1/2} N)$, where $\hat \theta_{\rm gec, \ell}^{(\rm ce)} = \sum_{i \in U} \tilde{\bm z}_i^\T \tilde{\bm \gamma}_{\rm ce} + \sum_{i \in A} d_i (y_i - \tilde{\bm z}_i^\T \tilde{\bm \gamma}_{\rm ce})$, $\hat \theta_{\rm gec, \ell}^{(\rm g)} = \sum_{i \in U} \tilde{\bm z}_i^\T \tilde{\bm \gamma}_g + \sum_{i \in A} d_i (y_i - \tilde{\bm z}_i^\T \tilde{\bm \gamma}_g)$, $\tilde{\bm \gamma}_{\rm ce} = \big\{ \sum_{i \in U} \frac{\pi_i \tilde{\bm z}_i \tilde{\bm z}_i^{\T}}{g_{\rm ce}^\prime(d_i)} \big\}^{-1} \sum_{i \in U} \frac{\pi_i \tilde{\bm z}_i y_i}{g_{\rm ce}^\prime(d_i)} $ and $\tilde{\bm \gamma}_g = \big\{ \sum_{i \in U} \frac{\pi_i \tilde{\bm z}_i \tilde{\bm z}_i^{\T}}{g^\prime(d_i)} \big\}^{-1} \sum_{i \in U} \frac{\pi_i \tilde{\bm z}_i y_i}{g^\prime(d_i)} $.
Furthermore, for Poisson sampling where $\{\delta_i\}_{i \in U}$ are independent, 
we have ${\mathbb V}(\hat \theta_{\rm gec, \ell}^{(\rm ce)}) \leq  {\mathbb V}(\hat \theta_{\rm gec, \ell}^{(\rm g)})$ for any entropy function $G(\cdot)$.
\end{proposition}

Proposition \ref{corollary2} shows that the GEC estimator using the contrast entropy is more efficient than the GEC estimator using other entropy functions when they are given the same set of calibration functions. The following proposition further compares the proposed contrast entropy estimator $\hat \theta_{\rm ce} = \sum_{i \in A} \hat{\omega}_{{\rm ce}, i}y_i$ in (\ref{eq:GEL}) to the augmented DS estimator $\hat{\theta}_{\rm ads}$ with the additional constraint in (\ref{eq:GEL-IBC}). Note that $\hat \theta_{\rm ce}$ and $\hat{\theta}_{\rm ads}$ use the same set of constraints for a fair comparison.
From (\ref{pel2-augment}), we have $\hat \theta_{\rm ads} = \hat \theta_{\rm ads, \ell} + o_p (n_0^{-1/2} N)$ where 
$\hat \theta_{\rm ads, \ell} = 
\sum_{i \in U} {\bm z}_i^\T {\bm \gamma}_g^{(\rm ads)} + \sum_{i \in A} d_i  (y_i - {\bm z}_i^\T {\bm \gamma}_g^{(\rm ads)} )$, $\bm z_i = (\bm x_i^{\T}, g_{\rm ce}(d_i))^{\T}$ and ${\bm \gamma}_g^{(\rm ads)} = \big(\sum_{i \in U} \bm z_i \bm z_i^\T\big)^{-1} \big(\sum_{i \in U} \bm z_i y_i\big)$.

\begin{proposition}
\label{corollary3}
Under the conditions of Theorem \ref{MainThm} and $n_0$ and $N$ being at the same order, we have $\hat \theta_{\rm ce} = \hat \theta_{\rm ce, \ell} + o_p(n_0^{-1/2} N)$, where $\hat \theta_{\rm ce, \ell} = \sum_{i \in U} {\bm z}_i^\T {\bm \gamma}_{\rm opt} + \sum_{i \in A} d_i (y_i - {\bm z}_i^\T {\bm \gamma}_{\rm opt})$, and ${\bm \gamma}_{\rm opt} = \big( \sum_{i \in U} (d_i - 1) {\bm z}_i {\bm z}_i^{\T} \big)^{-1} \sum_{i \in U} (d_i - 1) {\bm z}_i y_i$.
Furthermore, for Poisson sampling where $\{\delta_i\}_{i \in U}$ are independent, 
we have ${\mathbb V}(\hat \theta_{\rm ce, \ell}) \leq  {\mathbb V}(\hat \theta_{\rm ads, \ell})$.
\end{proposition}

The efficiency gain of the GEC estimator using the contrast entropy over the corresponding augmented DS estimator is due to the orthogonal projection of $\sum_i \delta_i d_i y_i$ onto $\sum_{i} (\delta_i d_i - 1) {\bm x}_i$ induced by the contrast entropy under Poisson sampling. 
This property may not hold for general entropy functions and sampling designs, as the variances of bias-corrected prediction estimators also depend on the outcome model. See Corollary \ref{corollary1} in Section \ref{ch4sec4} for a case where the GEC and DS estimators are asymptotically equivalent. 
It can be similarly shown that the same conclusions of Propositions \ref{corollary2} and \ref{corollary3} hold for the scaled weights when $n_0$ is much smaller than $N$.

\section{Unknown population-level inclusion probabilities}
\label{ch4sec7}

To apply the proposed method, the population total 
$ \sum_{i \in U} g(d_i)$ must be known in the debiasing constraint in (\ref{dcc}), which is possible if $\{\pi_i\}$ are known throughout the finite population. If $\{\pi_i\}$ is not available outside the sample, we cannot directly impose the constraint in (\ref{dcc}).
In this section, we consider the situation where $\sum_{i \in U} g(d_i)$ is unknown and modify the proposed method to estimate the target parameter $\theta_{\rm N}$ from the current survey.

Under this case, we modify the objective function in (\ref{prob}) as
\begin{equation}
\hat {\bm \omega}_{\rm m} = \argmin_{{\bm \omega}} \sum_{i \in A} \{G(\omega_i) - G(d_i) - g(d_i)(\omega_i - d_i)\} 
\mbox{ \ subject to \ } \sum_{i \in A} \omega_i \bm x_i = \sum_{i \in U} \bm x_i,
\label{kaa}\end{equation}
and propose the modified GEC estimator $\hat \theta_{\rm mgec} = \sum_{i \in A}\hat \omega_{{\rm m}, i} y_i$. Due to the convexity of $G(\omega)$, the objective function in (\ref{kaa}) is strictly convex with respect to $\bm \omega$.
Using the Lagrangian multiplier method, the solution $\hat {\bm \omega}_{\rm m}^{(1)}$ to (\ref{kaa}) satisfies
$\hat \omega_{{\rm m}, i} = f(\hat{\bm \lambda}_1^{\T} \bm x_i + g(d_i))$ for $i \in A$,
which is similar to the solution $\hat \omega_i$ of (\ref{prob}) in (\ref{fwgt}), but $\hat{ \lambda}_2$ is set to be 1. 
Note that $$G(\omega_i) - G(d_i) - g(d_i)(\omega_i - d_i) \approx g^{\prime}(d_i)(\omega_i - d_i)^2 / 2 = d_i^2 g^{\prime}(d_i)(\omega_i / d_i - 1)^2 / 2$$
for $\omega_i = d_i + o(1)$, which is a weighted $L_2$ distance between $\omega_i$ and $d_i$ with a weight depending on $G(\omega)$. 
While, for the DS method that minimizes the distance $D(\bm \omega, \bm d) = \sum_{i \in A}d_i G(\omega_i / d_i)$ with $G^{\prime}(1) = 0$, we have 
$d_i G(\omega_i / d_i) \approx d_i G^{\prime\prime}(1)(\omega_i / d_i - 1)^2 / 2$ 
for $\omega_i = d_i + o(1)$, which is also an $L_2$ distance between $\omega_i$ and $d_i$. However, the weight of the DS method is not dependent on the entropy function.
The following theorem provides the asymptotic expansion of the modified GEC estimator.

\begin{theorem}
\label{MainThm-modified}
Under the conditions of Theorem \ref{MainThm} and $n_0$ and $N$ being at the same order, the solution $\hat {\bm \omega}_{\rm m}$ to (\ref{kaa}) exists and is unique with probability approaching to 1.
Furthermore, the modified GEC estimator $\hat \theta_{\rm mgec}= \sum_{i \in A} \hat{\omega}_{\rm m, i} y_i$ satisfies $\hat \theta_{\rm mgec}  =  \hat \theta_{\rm mgec, \ell} + o_p(n_0^{-1/2} N)$, where 
\begin{equation*}
    \hat \theta_{\rm mgec, \ell}  =\sum_{i \in U} {\bm x}_i^\T \bm \beta_g + \sum_{i \in A} d_i (y_i - {\bm x}_i^\T \bm \beta_g) 
\mbox{ \ and \ }
\bm \beta_g = \bigg\{ \sum_{i \in U} \frac{\pi_i \bm x_i \bm x_i^{\T}}{g^\prime(d_i)} \bigg\}^{-1} \sum_{i \in U} \frac{\pi_i \bm x_i y_i}{g^\prime(d_i)}.    
    \end{equation*}
\end{theorem}

Note that $\bm \beta_g$ can be estimated by $\hat{\bm \beta}_{\rm g} = \left\{ \sum_{i \in A} \bm x_i \bm x_i^\T/ g'(d_i) \right\}^{-1} \sum_{i \in A} \bm x_i y_i/ g'(d_i)$, and the variance of the modified GEC estimator $\hat \theta_{\rm mgec}$ can be estimated similarly as (\ref{varhat}). Compared to the DS estimator $\hat{\theta}_{\rm  ds}$ with the regression coefficient ${\bm \beta}_N$, as discussed in (\ref{pel2}) and (\ref{pel2-augment}), our approach allows for different weights $\{g'(d_i)\}^{-1}$ in the regression coefficient $\bm \beta_{\rm g}$ using different entropies.
Similar as Corollary \ref{corollary1}, if $\pi_i$ is only determined by $\bx_i$ and there is a superpopulation linear regression model $y_i = \bx_i^{\T} \bbeta_0 + e_i$ for $y_i$ with $\mathbb{E}(e_i) = 0$ and $e_i \perp \bx_i$, then $\hat \theta_{\rm mgec}$ and $\hat{\theta}_{\rm  ds}$ are asymptotically equivalent. 
However, if the sampling mechanism is informative or the working regression model is misspecified, we can choose the contrast entropy to gain higher efficiency for the proposed modified GEC estimator. This is similar to the conclusion in Proposition \ref{corollary3}.
Recall that $\hat \theta_{\rm ds} = \hat \theta_{\rm ds, \ell} + o_p (n_0^{-1/2} N)$ from (\ref{pel2}). Let $\hat \theta_{\rm mce}$ denote the modified GEC estimator using the contrast entropy $G_{\rm ce}(\omega) = (\omega - 1) \log(\omega - 1) - \omega \log (\omega)$ in (\ref{kaa}). 

\begin{proposition}
\label{corollary4}
Under the conditions of Theorem \ref{MainThm}, Poisson sampling and $n_0$ and $N$ being at the same order, we have $\hat \theta_{\rm mce} = \hat \theta_{\rm mce, \ell} + o_p(n_0^{-1/2} N)$ and ${\mathbb V}(\hat \theta_{\rm mce, \ell}) \leq  {\mathbb V}(\hat \theta_{\rm ds, \ell})$, where $\hat \theta_{\rm mce, \ell} = \sum_{i \in U} {\bm x}_i^\T {\bm \beta}_{\rm opt} + \sum_{i \in A} d_i (y_i - {\bm x}_i^\T {\bm \beta}_{\rm opt})$ and ${\bm \beta}_{\rm opt} = \big( \sum_{i \in U} (d_i - 1) {\bm x}_i {\bm x}_i^{\T} \big)^{-1} \sum_{i \in U} (d_i - 1) {\bm x}_i y_i$.
\end{proposition}

This proposition shows the superiority of the proposed method using the contrast entropy over the DS estimation. 
Theorem \ref{MainThm-modified} and Proposition \ref{corollary4} can be similarly proven for the scaled weights when $n_0$ is much smaller than $N$.

\section{Simulation study}
\label{ch4sec8}
To test our theory, we performed a limited simulation study. We consider a finite population of size $N = 10,000$. A vector of two auxiliary variables $\bm x_i = (x_{1i}, x_{2i})^\T$ is available for $i = 1, \cdots, N$, where $x_{1i} \sim N(2, 1)$ follows the normal distribution with mean $2$ and standard deviation $1$, and $x_{2i} \sim {\rm Unif}(0, 4)$, uniform distribution in $[0, 4]$. We consider two super-population models to generate the study variable $y$: $y_{i} = x_{1i} + x_{2i} + e_{i}$ (Model 1) and $y_{i} = x_{1i} / 3 + x_{2i} / 3 + x_{1i} x_{2i}^2 / 4 + e_{i}$ (Model 2), where $e_{i} \sim N(0, 1)$, independent of $\bm x_i$.  From each of the finite populations, samples are selected using Poisson sampling with inclusion probability $\pi_i = \min (\Phi_3(-x_{1i} /2 -x_{2i} / 2 -2), 0.7)$, where $\Phi_3(\cdot)$ is the cumulative distribution function of the $t$ distribution with degree of freedom 3. 
The distribution of the design weights is right-skewed, resulting in some design weights being extremely large, as illustrated in Figure \ref{hist_boxplot}. The expected sample size is $\mathbb E(n) \approx 939$.
The model $R^2$ and the partial correlation between $\pi_i$ and $y_i$ after regressing out $\bm x_i$ were $R^2 = 0.6973$ and $\mbox{pCor}(\pi, y \mid \bm x) = -0.0040$, respectively, in Model 1. In Model 2,  $R^2 =  0.7893$ and $\mbox{pCor}(\pi, y \mid \bm x) = 0.6183$. 
For a fixed realization of the population, samples are generated repeatedly 1,000 times. We are interested in estimating the population mean $\mu_{\rm y} = N^{-1}\sum_{i \in U}y_{i}$ from the sampled data. 
We compare two scenarios: the population total  $ \sum_{i \in U} g(d_i)$ is available (Scenario 1) and $\sum_{i \in U} g(d_i)$ is not available (Scenario 2). From each sample, we compare the following estimators: 

\begin{figure}
    \centering
    \includegraphics[width=0.8  \linewidth]{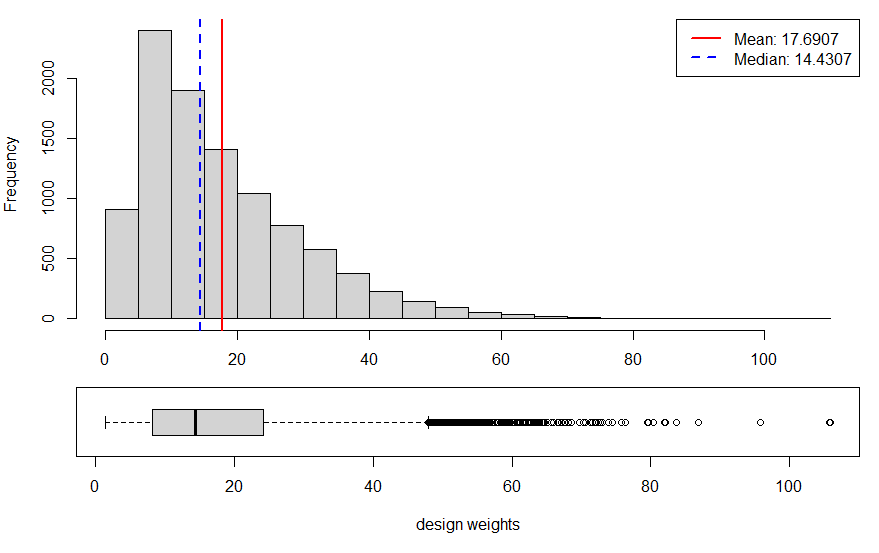}
    \caption{Histogram and boxplot of the design weights. The red solid line is the mean of the design weights and the blue dashed line is the median of the design weights.}
    \label{hist_boxplot}
\end{figure}

\begin{itemize}
    \item [\textbf{H\'ajek}] H\'ajek estimator: $\hat \mu_{\rm y, HT} = \del{\sum_{i \in A} d_i}^{-1}\del{\sum_{i \in A} d_iy_i}$.
    \item [\textbf{DS}] \cite{deville1992}'s divergence calibration estimator: $\hat \mu_{\rm y,  ds} = N^{-1}\sum_{i \in A}\hat w_i y_i$, where the calibration weight $\hat {\bm w} = \set{\hat w_i}_{i \in A}$ minimizes the divergence measure $D(\bm w, \bm d) = \sum_{i \in A}d_iG(\omega_i / d_i)$ defined in \eqref{DS} subject to $\sum_{i \in A} w_i = N$ and $\sum_{i \in A}w_i \bm x_i = \sum_{i \in U} \bm x_i$. Under Scenario 1 when $\sum_{i \in U} g(d_i)$ is known, an additional calibration constraint $\sum_{i \in A}\omega_i g(d_i) = \sum_{i \in U} g(d_i)$ is imposed for a fair comparison with the proposed estimator.
    
    \item [\textbf{GEC}] The proposed generalized entropy calibration estimator: $\hat \mu_{\rm y, cal} = N^{-1}\sum_{i \in A} \hat w_i y_i$, where the calibration weight $\hat {\bm w} = \set{\hat w_i}_{i \in A}$ maximizes the entropy $H(\bm \omega) = -\sum_{i \in A}G(w_i)$ subject to $\sum_{i \in A} w_i = N$, $\sum_{i \in A}w_i \bm x_i = \sum_{i \in U} \bm x_i$ and $\sum_{i \in A}w_i g(d_i) = \sum_{i \in U}g(d_i)$. Under Scenario 2, where $\sum_{i \in U} g(d_i)$ is unknown, the modified GEC method described in Section 6 is used.
\end{itemize}
For each of \textbf{DS} and \textbf{GEC} estimators, we consider the following entropy (divergence) functions $G(\omega)$: empirical likelihood ({\bf EL}) $G(\omega) = -\log \omega$, exponential tilting ({\bf ET}) $G(\omega) = \omega \log \omega - \omega$, contrast entropy ({\bf CE}) $G(\omega) = (\omega - 1)\log \del{\omega - 1} - \omega \log \omega$, and Hellinger distance ({\bf HD}) $G(\omega) = -4\sqrt{\omega}$.

Table \ref{tab2} presents the biases, standard errors (SE), and root mean squared errors (RMSE) of the estimators, along with the coverage rates (CR\%) of their 95\% confidence intervals, calculated from 1,000 Monte Carlo samples. When the sampling design is non-informative and the underlying working model of the calibration covariates is correct (Model 1), the \textbf{DS} estimator performs comparably to, or slightly better than, the \textbf{GEC} estimators in terms of RMSE. In addition, the estimators in Scenario 1 are also comparable to those in Scenario 2 as the additional debiasing covariate is unnecessary for predicting $y$ when the true superpopulation model  $y_i$ is linear in the calibration covariates $x_{1i}$ and $x_{2i}$.
On the other hand, when the underlying working model for calibration is incorrect, we observe different results. Under Model 2, the estimators in Scenario 1 show greater efficiency gain than the estimators in Scenario 2, as the augmented regression model can have a better prediction power than the reduced model without $g(d_i)$ in the covariates. Importantly, the \textbf{GEC} estimators are more efficient than the \textbf{DS} estimators (augmented \textbf{DS} estimators under Scenario 1) due to the efficiency gain in computing the regression coefficients. Specifically, the efficiency gain is outstanding under \textbf{CE}, which is consistent with our theory.
For the \textbf{ET} divergence, the performance of \textbf{DS} and \textbf{GEC} is essentially identical, which can be attributed to their asymptotic equivalence.



Throughout Table \ref{tab2}, the bias of all estimators is negligible, with the contribution of the squared bias to the mean squared error being less than 1\%.
Generally, coverage rates are close to the nominal 95\% level, within the bounds of experimental error. Under Model 2, coverage rates of the confidence intervals are slightly lower than the nominal coverage rates. This phenomenon aligns with the discussion by \cite{rao2003} on the undercoverage property of the regression estimator under model misspecification.

\spacingset{1.7}
\begin{table}
\centering
\begin{tabular}{@{}llccccccccc@{}}
\toprule
\multicolumn{2}{l}{}       & \multicolumn{4}{c}{Model 1} & & \multicolumn{4}{c}{Model 2} \\ \midrule
\multicolumn{2}{l}{}       & Bias   & SE    & RMSE  & CR & & Bias   & SE    & RMSE  & CR \\ \midrule
\multicolumn{2}{l}{H\'ajek}  & -0.30  & 8.18  & 8.18  & 96 & & -0.66  & 19.91 & 19.92 & 94 \\ \midrule
\multicolumn{11}{l}{Scenario 1 ($\sum_{i \in U} g(d_i)$ is known)} \\
\multirow{2}{*}{EL} & DS   & 0.04   & 4.01  & 4.01  & 96 & & -0.08  & 5.80  & 5.80  & 95 \\
                    & GEC  & 0.05   & 4.04  & 4.04  & 96 & & 0.02   & 5.36  & \bf 5.36  & 95 \\ \hdashline
\multirow{2}{*}{ET} & DS   & 0.03   & 4.02  & 4.02  & 96 & & 0.26   & 5.14  & \bf 5.14  & 94 \\
                    & GEC  & 0.03   & 4.02  & 4.02  & 96 & & 0.26   & 5.14  & \bf 5.14  & 94 \\ \hdashline
\multirow{2}{*}{CE} & DS   & 0.03   & 4.01  & 4.01  & 96 & & -0.14  & 6.10  & 6.10  & 95 \\
                    & GEC  & 0.05   & 4.04  & 4.04  & 96 & & -0.00  & 5.49  & \bf 5.49  & 94 \\ \hdashline
\multirow{2}{*}{HD} & DS   & 0.04   & 4.02  & 4.02  & 96 & & -0.01  & 5.36  & 5.36  & 95 \\
                    & GEC  & 0.04   & 4.02  & 4.02  & 96 & & 0.03   & 5.18  & \bf 5.18  & 95  \\
\midrule
\multicolumn{11}{l}{Scenario 2 ($\sum_{i \in U} g(d_i)$ is unknown)} \\
\multirow{2}{*}{EL} & DS  & 0.04   & 4.01  & 4.01  & 96 & & -0.11  & 7.89  & 7.89  & 94 \\
                    & GEC & 0.05   & 4.02  & 4.02  & 96 & & 0.04   & 6.63  & \bf 6.63  & 94 \\ \hdashline
\multirow{2}{*}{ET} & DS  & 0.04   & 4.01  & 4.01  & 96 & & -0.20  & 7.90  & \bf 7.90  & 94 \\
                    & GEC & 0.04   & 4.01  & 4.01  & 96 & & -0.20  & 7.90  & \bf 7.90  & 94 \\ \hdashline
\multirow{2}{*}{CE} & DS  & 0.04   & 4.01  & 4.01  & 96 & & -0.28  & 7.91  & 7.91  & 94 \\
                    & GEC & 0.05   & 4.03  & 4.03  & 96 & & 0.05   & 6.60  & \bf 6.60  & 94 \\ \hdashline
\multirow{2}{*}{HD} & DS  & 0.04   & 4.01  & 4.01  & 96 & & -0.16  & 7.89  & 7.89  & 94 \\
                    & GEC & 0.04   & 4.01  & 4.01  & 96 & & -0.09  & 7.03  & \bf 7.03  & 94 \\ \bottomrule
\end{tabular}
\caption{Bias $(\times 100)$, standard error (SE, $\times 100$), and root mean squared error (RMSE, $\times 100$) of the estimators, and coverage rate (CR, \%) of their 95\% confidence intervals under Model 1 (correct model) and Model 2 (incorrect model).}
\label{tab2}
\end{table}
\spacingset{1.9}

\section{Real data analysis}
\label{ch4sec9}
We present an application of the proposed method using data from a proprietary pesticide usage survey collected from GfK Kynetec in 2020. 
One of the main objectives of this survey is to estimate the total amount (\$) spent by farm operations on pesticides in each state of the United States. 

The survey was carried out by stratified sampling; the population was stratified by three factors: 50 states, 60 crops, and the size of a farm (integers from 1 to 7). Since larger farms tend to use greater amounts of pesticides, the sampling design assigned a greater proportion of the sample to larger farms within each stratum to reduce variance. See \cite{thelin2013estimation} for further details on the survey design. In order to handle item-nonresponse, the initial design weight was adjusted before calibration so that the weighted totals of the number of farms, stratified by crop type and farm size, align with the external benchmarks from the USDA Census of Agriculture. Despite this adjustment, we assume in this paper that all the sampled units are fully observed and the design weights are given, in order to focus solely on weight calibration under the assumption of no item nonresponse. 

For each farm $i$, the study variable $y_i$ is the dollar amount spent on the pesticide, including the herbicide, insecticide, and fungicide produced by the five largest agrochemical companies: BASF, Bayer, Corteva Agriscience, FMC, and Syngenta. The auxiliary variables $\bm x_i$ are the harvested areas (in acres) for each crop in each multi-county area, referred to as the Crop Reporting District (CRD). Total acres harvested for each crop-by-CRD combination are available from the USDA Census of Agriculture. Throughout the United States, there were more than 20,000 samples with more than 1,000 strata. We only report the results for four states for brevity.


For estimation, we compared the Horvitz-Thompson estimator (\textbf{HT}), generalized regression estimator (\textbf{Reg}) in \eqref{greg}, pseudo-empirical estimator (\textbf{PEL}) in \eqref{pel}, and the modified GEC estimator using empirical likelihood (\textbf{EL}), contrast-entropy (\textbf{CE}), and Hellinger distance (\textbf{HD}) described in Section 6. Since the design weights and the auxiliary variables are not available in each population unit, $\sum_{i \in U} g(d_i)$ is unknown in this dataset.

Table \ref{tabkynetec} in the supplementary material(SM) summarizes the point estimates, standard errors, and 95 \% confidence intervals of the estimators. All the calibration methods converge well and produce weights even when the number of auxiliary variables is greater than 20 as in Iowa. Incorporating auxiliary variables as in \textbf{Reg}, \textbf{PEL}, \textbf{EL}, \textbf{CE}, or \textbf{HD} improved performance compared to the Horvitz-Thompson estimator \textbf{HT}. The standard error of the proposed entropy calibration estimators using \textbf{EL} or \textbf{CE} was the smallest for all states reported. 

\section{Concluding remarks}
\label{ch4sec10} 

This paper introduces a novel framework for calibration estimation in survey sampling, leveraging generalized entropy as the objective function while incorporating design weights through a dedicated debiasing constraint to ensure design consistency. The proposed approach fundamentally differs from traditional methods by shifting design weights from the objective function to a constraint. The proposed calibration method implicitly utilizes an augmented regression model where the derivative of the entropy function, $g(d_i)$, acts as an additional covariate. This not only achieves design consistency but also offers potential efficiency improvements, particularly when the standard working regression model is misspecified, as demonstrated in our simulation study.

The resulting calibration weights can be applied to multiple outcome variables, yielding design-consistent estimators irrespective of the specific outcome variable. The efficiency gain for any particular outcome variable hinges on the predictive power of the additional covariate $g(d_i)$. If the coefficient associated with $g(d_i)$ in the augmented regression model is significant, the proposed estimator is likely to be more efficient than traditional calibration estimators. Conversely, if this coefficient is insignificant, incorporating the constraint might slightly increase variance. While a preliminary significance test for the coefficient of $g(d_i)$ could guide the decision to use the augmented model, the potential efficiency gains when the working model is misspecified are asymptotically of a higher order than the potential variance increase when the model is correct. Thus, asymptotically, the advantages often outweigh the risks.

We have also identified the contrast-entropy function as yielding an asymptotically optimal estimator under Poisson sampling in a pairwise comparison framework when compared to other entropy functions using the same set of calibration constraints. However, this theoretical optimality relative to the specific constraints might not always translate to the most efficient estimate in practice when comparing estimators derived from different entropy functions, as their calibration functions are different.
Practitioners might consider fitting the augmented regression models corresponding to different candidate entropy functions and selecting the one that maximizes a goodness-of-fit measure like $R^2$ for the specific outcome variable(s) of interest. 

While the calibration weighting has been developed for probability samples, the calibration weighting is also used for non-probability samples. In this case, the main goal for calibration is to reduce the selection bias in the non-probability samples \citep{kott2012providing}. 
The proposed method can be directly applied in this case, but it is beyond the focus of the paper and will be presented elsewhere.  
In practice,  unit nonresponse or undercoverage often necessitates adjustments to design weights. 
In such cases, the generalized entropy function can be chosen such that the first-order derivative of its convex conjugate function equals the inverse of the response propensity function \citep{slud2022}.
Also, when $p=\mbox{dim} ( \bx)$ is large, we can apply generalized entropy with soft calibration using the $L_2$ norm \citep{guggemos2010} or the $L_1$ norm \citep{mc2017}. Once the debiasing constraint is satisfied, other benchmarking constraints can be relaxed to accommodate high-dimensional auxiliary variables.  
In addition, an \texttt{R} package (\texttt{GECal}) implementing the proposed debiasing calibration weighting is available on \texttt{CRAN}.



\bibliographystyle{apalike}  
\bibliography{reference}

\end{document}